\documentstyle[pre,eqsecnum,aps]{revtex}
\input epsf
\begin{document}

\title{Tricritical universality in a two-dimensional spin fluid}

\author{N. B. Wilding and P. Nielaba}
\address{Institut f\"{u}r Physik, Johannes Gutenberg Universit\"{a}t, \\
Staudinger Weg 7, D-55099 Mainz, Germany.}

\date{May 1995}
\setcounter{page}{0}
\maketitle

\begin{abstract}

Monte Carlo simulations are used to investigate the tricritical point
properties of a 2d spin fluid.  Measurements of the scaling operator
distributions are employed in conjunction with a finite-size scaling
analysis to locate the tricritical point and determine the directions
of the relevant scaling fields and their associated tricritical
exponents.  The scaling operator distributions and exponents are shown
to match quantitatively those of the 2d Blume-Capel model, confirming
that both models belong to the same universality class. Mean-field
calculations of the tricritical point properties are also compared
with the simulation measurements.

\end{abstract}
\thispagestyle{empty}
\begin{center}
PACS numbers 64.60Fr, 64.70.Fx, 05.70.Jk
\end{center}
\newpage

\section{Introduction}
\label{sec:intro}

For tricritical phenomena, the highest dimension in which
non-classical behaviour can be observed is $d=2$
\cite{LAWRIE}. Consequently, 2d tricritical phenomena has been the
subject of a large number of previous investigations, employing a wide
variety of techniques, including series expansions \cite{SAUL},
mean-field theory \cite{FURMAN}, renormalisation group (RG)
\cite{OLIVEIRA,BAKCHICH,BURKHARDT,BERKER,YEOMANS}, transfer matrix
\cite{BEALE,HERRMANN,RIKVOLD,ALCARAZ}, Monte Carlo simulations
\cite{AYAT,KIMEL} and Monte Carlo RG methods
\cite{LANDAU,LANDAU1,HONDA}. To date, however, this interest has
focused almost exclusively on lattice-based spin models such as the
Blume-Capel model or the spin-$\frac{1}{2}$ next-nearest-neighbour
Ising model. In this paper, we report the first detailed simulation
study of 2d tricritical behaviour in an off-lattice spin fluid model.

The model we consider is a simplified representation for a liquid of
two-state molecules, and has been the subject of a number of previous
studies in both its classical and quantum regimes \cite{STRATT}. In
the present work, however, we shall consider only the classical limit,
for which the configurational energy is given by:

\begin{equation}
\Phi(\{\vec{r},s\})=-J(r_{ij})\sum_{i<j}^Ns_is_j+\sum_{i<j}^NU(r_{ij})
\label{eq:pot}
\end{equation}
with $s_i=\pm 1$ and where $U(r_{ij})$ is chosen to be a hard disk potential
with diameter $\sigma$. The distance-dependent spin coupling
parameter $J(r_{ij})$ is assigned a square well form:

\begin{eqnarray}
J(r)=J \hspace{1cm} &\sigma < r < 1.5\sigma \nonumber \\
J(r)=0 \hspace{1cm} &{\rm elsewhere}
\end{eqnarray}

The phase diagram of this model is expected to correspond to the
situation depicted schematically in figure~\ref{fig:pdschem}. For high
temperatures, there exists a line of Ising critical points (the
so-called `critical line') separating a ferromagnetic fluid phase from
a paramagnetic fluid phase. The particle density varies continuously
across this line. As one follows the critical line to lower
temperatures, however, the size of the particle density fluctuations
grows progressively. Precisely at the tricritical point, the
fluctuations in both the particle density and magnetisation are
simultaneously divergent. Lowering the temperature still further
results in a phase separation between a low density paramagnetic gas
and a high density ferromagnetic liquid. For subtricritical
temperatures, the phase transition between these two phases is first
order.

Owing to the interplay between the density and magnetisation
fluctuations, the tricritical properties of the spin fluid system are
expected to differ qualitatively from those on the critical line.
General universality arguments \cite{KADANOFF} predict that for a
given spatial dimensionality, fluids with short-ranged interactions
should exhibit the same tricritical properties as lattice-based spin
systems.  However, since fluids possess a continuous translational
symmetry that lattice models do not, this proposal needs be
checked. Additionally, experience with `ordinary' (Ising) critical
behaviour in simple fluids such as the Lennard-Jones fluid
\cite{WILDING3,WILDING4}, shows that the reduced symmetry of fluids
can profoundly influence certain non-universal aspects of the critical
properties. Principal among these, are the directions of the relevant
scaling fields associated with the fixed point, and the distribution
functions of observables such as the particle density and energy. It
is thus of interest to assess the extent of these `field-mixing'
effects in the tricritical fluid and to compare it with the situation
at the liquid-vapour critical point of simple fluids.

An accurate determination of the universal forms of the tricritical
scaling operator distribution is also of considerable value. Such
distributions are {\em unique} to a universality class and hence
knowledge of their forms would be of considerable practical utility to
future simulation studies of 2d tricriticality, serving as they do to
simplify the computational task of locating the tricritical
parameters. Moreover, as we shall see, the forms of the scaling operator
distribution functions can impart important physical insight into the nature
of the tricritical fluctuations.

Our paper is broadly organised as follows. In section~\ref{sec:meth} we
describe the finite-size scaling methods and other computational
techniques employed in the study. We then proceed in
section~\ref{sec:res} to detail the application of these techniques to
Monte Carlo simulations of both the 2d spin fluid model described above,
and the 2d Blume-Capel. The simulations yield accurate estimates of the
location of the tricritical point for both models, as well as the
universal forms of the tricritical scaling operator distributions and
the directions of the relevant scaling fields. In the case of the spin
fluid model, the estimates for the tricritical point parameters are
compared with the results of a mean field calculation. Finally
section~\ref{sec:concs} details our conclusions.

\section{Background}
\label{sec:meth}

The techniques we employ in this work have been previously developed
in the context of simulation studies of Ising critical phenomena in
a variety of fluid models, including a decorated lattice gas model
\cite{WILDING1,WILDING2}, a lattice model for polymer mixtures
\cite{MUELLER}, and both the two and three-dimensional Lennard-Jones
fluids \cite{WILDING3,WILDING4}. In common with the strategy pursued
in these previous works, we have chosen to work within the grand
canonical ensemble, use of which affords effective treatment of the
particle density fluctuations which are a central feature of fluid
critical behaviour.

Let us assume our system to be contained in a volume $L^d$, with $d=2$
in the simulations to be described later.  The grand partition
function is given by

\begin{equation}
\label{eq:bigzdef}
{\cal Z}_L  = \sum _{N=0}^{\infty }\sum_{\{s_i\}}\prod _{i=1}^{N}
\left\{\int d\vec{r}_i\right\} e^{-\beta\left[\Phi (\{ \vec{r},s \} )
+ \mu N +H\sum_j^N s_j\right] }
\label{eq:pf}
\end{equation}
where $N$ is the particle number, $\beta=(K_BT)^{-1}$ is the inverse
temperature, $\mu$ is the chemical potential and $H$ is the uniform
applied magnetic field.

The observables of chief concern to the present study are the
(reduced) particle density

\begin{equation}
\rho=L^{-d}N\sigma^d ,
\end{equation}
the configurational energy density (which we express in units of $J$)
\begin{equation}
u=L^{-d}J^{-1}\Phi(\{{\vec{r},s}\}),
\end{equation}
and the magnetisation.
\begin{equation}
m=L^{-d}\sum_is_i
\end{equation}

The coarse-grained behaviour of the system in the vicinity of the
tricritical point is controlled by three relevant scaling fields
\cite{LAWRIE,GRIFF,WEGNER} which we denote $g,\lambda$ and $h^\prime$. In
general these scaling fields are each expected to comprise linear
combinations of the three thermodynamic fields $T, \mu$ and $H$
\cite{REHR,ALCARAZ}.  For the spin fluid model considered in this work,
however, the configurational energy is invariant with respect to sign
reversal of the spin degrees of freedom. This special symmetry implies
that the tricritical point lies in the symmetry plane $H=0$, and that
the scaling field $h^\prime$ coincides with the magnetic field $H$,
being orthogonal to the $\mu-T$ plane containing the other two scaling
fields, $g$ and $\lambda$. Thus we can write

\begin{mathletters}
\begin{eqnarray}
\label{eqn:scafldsa}
h^\prime & = & H-H_t \\
\lambda & = & (\mu-\mu_t)+r(T-T_t) \\
g & = & T-T_t+s(\mu-\mu_t)
\label{eqn:scafldsc}
\end{eqnarray}
\end{mathletters}
where the subscript $t$ signifies tricritical values and the
parameters $s$ and $r$ are system-specific  `field mixing' parameters
that control the directions of the scaling fields in the $\mu$--$T$
plane.  The scaling fields $g$ and $\lambda$ are depicted schematically in
figure~\ref{fig:pdschem}(b). One sees that $g$ is tangent
to the coexistence curve at the tricritical point \cite{GRIFF}, so that
the field mixing parameter $r$ may be identified simply as the limiting
tricritical gradient of the coexistence curve. The scaling field
$\lambda$, on the other hand, is permitted to take a general direction
in the $\mu$--$T$ plane which does not necessarily have to coincide with any
special direction of the phase diagram \cite{REHR}.

Conjugate to each of the scaling fields are scaling operators, defined
by the requirements
\begin{mathletters}
\begin{eqnarray}
\langle{\cal M}\rangle & \equiv & L^{-d}\partial \ln {\cal Z}_L/\partial
h^\prime \\
\langle{\cal D}\rangle & \equiv & L^{-d}\partial \ln {\cal Z}_L/\partial
\lambda \\
\langle{\cal E}\rangle & \equiv & L^{-d}\partial \ln {\cal Z}_L/\partial g
\end{eqnarray}
\end{mathletters}
from which it follows (utilising equations~\ref{eq:pf} and
\ref{eqn:scafldsa}---\ref{eqn:scafldsc}), that
\begin{mathletters}
\begin{eqnarray} {\cal M} & = & m \\
{\cal D} & = & \frac{1}{1-sr}[\rho-su ] \\
{\cal E} & = & \frac{1}{1-sr}[u-r\rho ]
\end{eqnarray}
\end{mathletters}

Motivated by our experience with ordinary critical phenomena in simple
fluids \cite{WILDING3} we make the following finite-size scaling {\em
ansatz}  \cite{PRIVMAN} for the limiting (large L) near-tricritical
distribution of
$p_L(\rho,u,m)$

\begin{equation}
p_L(\rho,u,m)\simeq\frac{1}{1-sr}\tilde{p}_L(a_1^{-1}L^{d-y_1}{\cal M},
a_2^{-1}L^{d-y_2}{\cal D},a_3^{-1}L^{d-y_3}{\cal
E},a_1L^{y_1}h^\prime,a_2L^{y_2}\lambda,a_3L^{y_3}g)
\label{eq:ansatz}
\end{equation}
where $\tilde{p}_L$ is a universal scaling function,  the $a_i$ are
non-universal metric factors and the $y_i$ are the standard
tricritical eigenvalue exponents \cite{NOTE1}.
Precisely at the tricritical point, the tricritical scaling fields
vanish identically and the last three arguments of
equation~\ref{eq:ansatz} can be simply dropped, yielding

\begin{equation}
p_L(\rho,u,m)\simeq\frac{1}{1-sr}\tilde{p}_L^\star(
a_1^{-1}L^{d-y_1}{\cal M},a_2^{-1}L^{d-y_2}{\cal D},a_3^{-1}L^{d-y_3}{\cal
E})
\label{eq:ptri}
\end{equation}
where $\tilde{p}_L^\star$ is a universal and scale invariant function
characterising the tricritical fixed point.

In what follows we shall explicitly test the proposed universality of
equation~\ref{eq:ptri} for the case of the spin fluid model, by
obtaining the form of $\tilde{p}_L^\star$ and comparing it with that for
the tricritical 2d Blume-Capel model.

\section{Results}
\label{sec:res}
\subsection{computational aspects}

The Monte-Carlo simulations of the spin fluid model were performed
using a Metropolis algorithm within the grand canonical ensemble.
Particle insertions and deletions were carried out using the
prescription of Adams \cite{AD3,ALLEN}. When attempting particle
insertions the spin of the candidate particle was randomly assigned
the value $+1$ or $-1$ with equal probability. Spin flip attempts were
performed at the same frequency as insertion/deletion attempts.

In order to facilitate efficient computation of interparticle
interactions, the periodic simulation space of volume $L^2$ was
partitioned into $l^2$ cubic cells each of side $1.5\sigma$,
corresponding to the interaction range of the interparticle potential
(cf. equation~\ref{eq:pot}). This strategy ensures that interactions
emanating from particles in a given cell extend at most to particles
in the $8$ neighbouring cells. We chose to study three system sizes
corresponding to $l=12,16$ and $20$, containing, at coexistence,
average particle numbers of $\langle N \rangle=120$, $210$ and $330$
respectively. For the $l=12,16$ and $20$ system sizes, equilibration
periods of $10^5$ Monte Carlo transfer attempts {\em per cell} (MCS)
were utilised, while for the $l=16$ and $l=20$ system sizes up to
$2\times10^6$ MCS were employed. Sampling frequencies ranged from $25$
MCS for the $l=12$ system to $100$ MCS for the $l=20$
system. Production runs amounted to $1\times10^7$ MCS for the $l=12$
and up to $5\times10^7$ MCS for the $l=20$ system size. At coexistence
the average acceptance rate for particle transfers was approximately
$16\%$, while for spin flip attempts the acceptance rate was
approximately $5\%$.

During the production runs, the joint probability distribution
$p_L(\rho,u,m)$ was obtained in the form of a histogram. In order to
increase computational efficiency, the histogram extrapolation technique
\cite{FERRENBERG} was employed. Use of this technique permits
histograms obtained at one set of model parameters to be reweighted to
yield estimates appropriate to another set of model parameters. The
method is particularly effective close to a critical point where, owing
to the large fluctuations, a single simulation permits extrapolation
over the entire critical region.

As an aid to locating the tricritical point, the cumulant intersection
method was employed \cite{BINDER2}. The fourth order cumulant ratio
$U_L$ is a quantity that characterises the form of a distribution
\cite{CRAMER}. It
is defined in terms of the fourth and second moments of
a given distribution

\begin{equation}
U_L=1-\frac{<m^4>}{3<m^2>^2}
\end{equation}
The tricritical scale invariance of the distributions $p_L({\cal D}),
p_L({\cal M})$ and $p_L({\cal E})$, (as expressed by
equation~\ref{eq:ptri}), implies that at the tricritical point (and
modulo corrections to scaling), the cumulant values for all system
sizes should be equal. The tricritical parameters can thus be found by
measuring $U_L$ for a number of temperatures and system sizes along
the first order line, according to the prescription given below.
Precisely at the tricritical temperature the curves of $U_L$
corresponding to the various system sizes are expected to intersect
one another at a single common point.

\subsection{The 2d Blume-Capel model}
\label{sec:BC}

In seeking to confirm the proposed universality linking the
tricritical point of the 2d spin fluid model to those of 2d lattice
models, it is first necessary to determine the tricritical operator
distribution functions for a simple lattice model. To this end we have
also performed a simulation study of the Blume-Capel model on a
periodic square lattice, the Hamiltonian of which is given by

\begin{equation}
{\cal H}=-J\sum_{<i,j>}s_is_j + D\sum_is_i^2 +H\sum_is_i
\end{equation}
with $s_i=-1,0,1$. Here $H$ is a uniform magnetic field and $D$ is the
so-called `crystal field'. As with the 2d spin fluid model, the
symmetry of the configurational energy under spin sign reversal implies that
the
tricritical point lies in the symmetry plane $H=0$.

Previous MCRG \cite{LANDAU,LANDAU1} investigations place the
tricritical point of the 2d Blume-Capel model at $K_BT_t/J=0.609(3),
D_t=1.965(15)$, while a more recent transfer matrix study \cite{BEALE} gives
$K_BT_t/J=0.610(5), D_t=1.965(5)$. Using these estimates as an initial
guide, we performed extensive Monte Carlo simulations of the model
using a vectorised Metropolis algorithm on a Cray-YMP. Four system
sizes of linear extent $L=12,20,32,40$ were studied and following
equilibration, runs ranging from $5\times10^6$ Monte Carlo sweeps
(MCS) for the L=12 system, to $2\times10^7$ MCS for the L=40 system
were performed. The quantities measured in the course of these runs
were

\begin{mathletters}
\begin{eqnarray}
\rho^\prime &=& \sum s_i^2 , \\
          u &=& -J\sum_{<i,j>}s_is_j , \\
          m &=& \sum s_i .
\end{eqnarray}
\end{mathletters}
Here we note that $\rho^\prime=\sum s_i^2$ plays the same r\^{o}le as
the density in the spin fluid model, being discontinuous across the
first order line but continuous on the critical line. This fact is most
clearly evident in the lattice gas representation of the Blume-Capel model,
where the crystal field $D$ appears as a chemical potential \cite{LAJZEROWICZ}.

During the simulations, the joint distribution $p_L(\rho^\prime,u,m)$
was collected in the form of a histogram. To determine the locus of
the first order line (and hence locate the tricritical point in which
it formally terminates), the coexistence symmetry criterion for the
operator distribution $p_L({\cal D})$ was utilised. This criterion is
the analogue for asymmetric first order transitions, of the
order-parameter distribution symmetry condition applicable to
symmetric first order transitions such as that of the subcritical
simple Ising model \cite{WILDING2}. For a given temperature $T$, the
first order transition point can thus be located by tuning the crystal
field $D$ and the value of the field mixing parameter $s$, within the
histogram reweighting scheme, until the operator distribution
$p_L({\cal D})$ is symmetric in ${\cal D}-\langle{\cal D}\rangle$.

The first order line and its finite-size analytical extension
\cite{WILDING3} was determined in this way for temperatures in the
range $K_BT/J=0.59$--$0.625$, and for each of the $4$ system
sizes. The corresponding values of the cumulant ratio $U^{\cal D}_L$
{\em along} this coexistence line are presented in
figure~\ref{fig:BC_UL} as a function of the temperature. To within
numerical uncertainties the cumulant values for each system size
intersect at a common temperature, which we estimate as
$K_BT_t/J=0.608(1)$. The corresponding estimate for the tricritical
field is $D_t=1.9665(3)$. Clearly these values are in excellent
agreement with the aforementioned estimates of previous studies.

In the following subsection we compare the measured forms of the
tricritical operator distributions $p_L({\cal M}),p_L({\cal D})$ and
$p_L({\cal E})$ of the 2d Blume-Capel, with those of the 2d spin fluid
model.

\subsection{The spin fluid model} \label{sec:sf}

The procedure for locating the tricritical point of the 2d spin fluid
model followed the same pattern as that for the 2d Blume-Capel model
described above, except that in the present case no prior estimates
for the tricritical point were available. It was thus necessary to
search for the approximate location of the tricritical point by
performing a number of short runs in which a temperature was chosen
and the chemical potential tuned. Observations of the behaviour of the
density from these short runs suggested that the tricritical point lay
close to the parameters $K_BT/J=0.58, \beta\mu=-1.915$.

Having obtained an approximate estimate of the location of the
tricritical point, long runs were carried out for each of the three
system sizes $l=12,16,20$. As with the Blume-Capel model, the
coexistence symmetry condition was then applied to the operator
distribution $p_L({\cal D})$ in conjunction with the histogram
reweighting scheme, in order to determine the first order line and its
analytic extension in the $\mu$--$T$ plane. The locus of this line is
shown in figure~\ref{fig:mu_T}, while the measured values of $U^{\cal
D}_L$ along this coexistence line are shown in figure~\ref{fig:fl_UL}
for the three system sizes $l=12,16,20$. Clearly the curves of
figure~\ref{fig:fl_UL} have a single well-defined intersection point,
from which we estimate the tricritical temperature as being
$K_BT_t/J=0.581(1)$. The associated estimate for the chemical
potential is $\beta\mu_t=-1.916(2)$.  Typical near-tricritical
configurations for the $l=20$ system are shown in figure

In figure~\ref{fig:collapse} we present the forms of the operator
distributions $p_L({\cal M})$, $p_L({\cal D})$ and $p_L({\cal E})$
corresponding to the designated values of the tricritical
parameters. The value of the field mixing parameter $r$ implicit in
the definition of ${\cal E}$, was assigned the value $r=-2.82$, as
obtained from the measured gradient of the phase boundary in the
$\mu$--$T$ plane at the tricritical point. The value of the field
mixing parameter $s$ was assigned, as previously described, so that
$p_L({\cal D})$ satisfied the symmetry condition. However, the
resulting estimates of $s$ were found to exhibit a systematic
finite-size dependence.  This effect has also been previously noted
(albeit with much reduced magnitude) in a recent study of critical
phenomena in the Lennard-Jones fluid, and is traceable to the
finite-size dependence of the average critical energy
\cite{WILDING2,WILDING3}. For the three system sizes, $l=12,16,20$, we
found $s=-0.031,-0.020,-0.013$ respectively. Interestingly, these
values are at least an order of magnitude smaller than those measured
at the critical point of the 2d and 3d Lennard-Jones fluid, a finding
which we discuss further in section~\ref{sec:concs}. This smallness
implies that the scaling field $\lambda$ almost coincides with the
$\mu$ axis of the phase diagram.

Also included in figure~\ref{fig:collapse} are the measured tricritical
operator distributions for the 2d Blume-Capel model. In accordance with
convention, all the operator distributions have been scaled to unit norm
and variance. Clearly in each instance and for each system size, the
operator distributions collapse extremely well onto one another as well
as onto those of the tricritical Blume-Capel model.

The measured scaling operator distributions also serve to furnish
estimates of the eigenvalue exponents $y_1,y_2$ and $y_3$
characterising the three relevant scaling fields. These exponents are
accessible via the respective finite-size scaling behaviour of
tricritical distributions of $p_L({\cal M})$, $p_L({\cal D})$ and
$p_L({\cal E})$.  Specifically, consideration of the scaling
form~\ref{eq:ptri} shows that the typical size of the fluctuations in
a given operator ${\cal O}$ vary with system size like $\delta{\cal
O}\sim L^{-(d-y_i)}$. Comparison of the standard deviation of a given
operator distributions as a function of system size thus affords
estimates of the appropriate exponents. From the measured variance of
the spin fluid operator distributions, we find $y_1=1.93(1),
y_2=1.80(1), y_3=1.03(7)$.  As far as $y_1$ and $y_2$ are concerned,
these estimates are in excellent agreement with exact conjectures
$y_1=77/40,y_2=9/5$ \cite{NEINHUIS,NIJS,PEARSON}.  The situation
regarding the eigenvalue exponent $y_3$, on the other hand, is not so
satisfactory, there being a sizable discrepancy with the exact value
of $y_3=4/5$.  This discrepancy stems, we believe, from two
sources. Firstly, since the operator distribution $p_L({\cal E})$ is
highly sensitive with respect to the designation of the value of the
field mixing parameter $r$ implicit in the definition of ${\cal E}$,
small uncertainties in the estimate of $r$ can lead to significant
larger errors in the measured variance of $p_L({\cal E})$. A similar
effect was also previously observed at the liquid-vapour critical
point of the Lennard-Jones fluid \cite{WILDING4}. Secondly, and as we
discuss in section~\ref{sec:concs}, the near Gaussian character of
$p_L({\cal E})$ signifies the absence of strong fluctuations in the
${\cal E}$, in which case it is questionable whether a finite-size
scaling can be reliable applied to $p_L({\cal E})$ at all.

In view of this problem we have adopted a rather different approach
for measuring $y_3$ based on the scaling properties of $U^{\cal D}_L$,
a quantity which does show strong fluctuations and which is also
insensitive to the designation of the field mixing parameter $r$. It
can be shown \cite{BINDER2}, that the maximum slope of the cumulant
ratio $\frac{dU_L}{dT}$ near $T_c$ varies with system size like
$L^{y_3}$. Using the histogram extrapolation technique, we have
obtained the temperature dependence of this slope for the spin fluid
model. The results yield the estimate $y_3=0.83(5)$, which agrees to
within error with the exact conjecture.

Turning now to the observables $m,\rho$ and $u$, we plot in
figure~\ref{fig:flbare} the measured distributions of these quantities
at the assigned tricritical parameters. Here we note that as is the
case with ordinary critical phenomena in the Lennard-Jones fluid
\cite{WILDING4}, the energy distribution $p_L(u)$ differs
qualitatively in form from the operator distribution $p_L({\cal E})$.
This finding, the origin of which is explained in detail in
reference~\cite{WILDING2}, reflects the coupling of the tricritical
energy fluctuations to the density, the latter of which are stronger
and thus dominate for large $L$.  The influence of this coupling is
also discernible as a small asymmetry in the tricritical density distribution.
For
the average tricritical density we find $\rho_t=0.374(1)$, while for
the average tricritical energy density we find $u_t=0.778(2)$. The
average magnetisation is of course strictly zero on symmetry grounds.

Finally in this subsection, table~\ref{tab:ULs} summarises the measured
values of the fourth order cumulant ratios $U_L^{\cal O}$ for each of
the three scaling operator distributions at the assigned values of the
tricritical parameters.

\subsubsection{Mean-field calculations}
\label{sec:meanfd}

In this section we describe a simple mean-field calculation of the
tricritical parameters of the 2d spin fluid.

Let $p(\rho,\mu;h)$ denote the pressure in the system for a given
chemical potential $\mu$ and volume $V$ in the thermodynamic limit, where
the equilibrium
density $\rho$ is given by the ratio of the average number of particles
$<N>$ and the volume $V$, $\rho = \lim_{V \to \infty} (<N>/V)$. Then
\begin{equation}
p(\rho,\mu;h) = min_{\rho'} \left[ -f(\rho';h) + \mu \rho' \right]
\label{MF1}
\end{equation}
where $f(\rho';h)$ is the free energy per volume,
\begin{equation}
f(\rho';h) = \lim_{V \to \infty} \frac{-1}{\beta V} \ln tr
\left[ \exp \left( -\beta H^{N} (h) \right) \right] \;\;.
\label{MF2}
\end{equation}
In the mean field approximation we assume an interaction between
a spin $s_1$ and an effective field $(q/N) \sum_{i>1} s_i =
q m$, where $q$ is the effective coordination number and the
$N$-particle Hamiltonian is written as
\begin{equation}
H^N_{MF} = \sum_{(i<j)} U(r_{ij}) -\sum_{i=1}^N \left[ q m(h) + h \right]
s_i
\label{MF2A}
\end{equation}
and the mean field free energy $f_{MF}$ is
\begin{equation}
f_{MF}(\rho';h) = \lim_{V \to \infty} \frac{-1}{\beta V} \ln tr
\left[ \exp \left( -\beta H_{MF}^{N} (h) \right) \right]
\label{MF2B}
\end{equation}
\begin{equation}
f_{MF}(\rho';h) = f_{cl}(\rho') +
min_m \left[ \frac{q \rho' m^2}{2} - \frac{\rho'}{\beta}
\ln 2 \cosh \left( \beta (q m + h) \right) \right]
\label{MF3}
\end{equation}
$f_{cl}(\rho')$ is the free energy of a classical system with
Hamiltonian $H_{cl} = \sum_{(i<j)} U(r_{ij})$
and the second term on the right hand side of Eq.~(\ref{MF3})
reaches its minimum at
\begin{equation}
m(h) = \tanh \left[\beta \left( q m(h) + h \right) \right]
\label{MF4}
\end{equation}
Since the coordination number $q$ in the fluid is not fixed we approximate
the effective field on one particle by
\begin{equation}
q m = m \rho' \int d^2r J(r) g(r)
\label{MF4A}
\end{equation}
where the fluid correlation function $g(r)$ is taken from the Percus
Yevick solution for hard discs, which can be found numerically.

Two phase coexistence between a gas phase at density $\rho_g$ and a
liquid phase phase at density $\rho_l$ is given by equal pressure in the
two phases

\begin{equation}
p(\rho_g,\mu;h) = p(\rho_l,\mu;h)
\label{MF5}
\end{equation}
This condition determines the mean--field chemical potential
$\mu_{MF}$ for phase coexistence via Eq.~(\ref{MF1}). Implementing
this criterion numerically, we find $T_t^{mf}=1.006$
$\beta\mu_t^{mf}=-1.319$. Clearly this estimate for $T_t$ seriously
overestimates the measured tricritical temperature ($T_t=0.581$)
showing that mean field calculations of this type cannot be relied upon to
provide
accurate tricritical data, at least for 2d systems.

\section{Conclusions}
\label{sec:concs}

In summary we have demonstrated that the tricritical ordering operator
distributions of the 2d spin fluid can be mapped into excellent
correspondence with those of the 2d Blume-Capel model. The existence of
such a mapping represents perhaps the most stringent test of
universality. There can thus be little doubt that despite their very
different microscopic character, the two systems do indeed share a
common fixed point.

With regard to the scaling operator distribution themselves, we note
that the form of $p^\star_L({\cal E})$ is (to within the precision of
our measurements) essentially Gaussian, as evidenced by the very small
value of the cumulant ratio $U_L^{\cal E}=0.003(3)$.  This Guassian
behaviour implies that the tricritical fluctuations in ${\cal E}$ are
extremely weak and is thus a consequence of the central limit theorem.
This weakness is further manifest in the very small measured values of
the field mixing parameter $s$, as well as in the near absence of
asymmetry in the tricritical density distributions, a situation which
contrast markedly with that of the 2d and 3d Lennard-Jones fluids,
where much stronger field mixing effects are observed in the density
distribution \cite{WILDING4}.

Finally, the tricritical form of $\tilde{p}_L^\star({\cal M})$ merits special
comment. We note that the {\em three-peaked} form of this distribution
differs radically from the universal magnetisation distribution of the
critical Ising model, which is strongly {\em double}-peaked in two
dimensions \cite{BINDER2,NICOLAIDES}. The existence of a three peaked
structure for tricritical phenomena reflects the additional coupling
that arises between the magnetisation and the density
fluctuations. Specifically, the central peak corresponds to
fluctuation to small density, which are concomitant with an overall
reduction in the magnitude of the magnetisation
(cf. figure~\ref{fig:config}). Were one, however, to depart from the
tricritical point along the critical line, these density fluctuations
would gradually die out and a crossover to a magnetisation
distribution having the double-peaked Ising form would be expected. In
future work we intend to investigate the nature of this crossover in
detail.

\subsection*{Acknowledgements}

A generous allocation of Cray-YMP computer time at the RHRK,
Universit\"{a}t Kaiserslautern and HLRZ, J\"{u}lich is gratefully
acknowledged. The authors have benefitted from helpful discussions
with K. Binder, B.  D\"{u}nweg, K.K. Mon and M. M\"{u}ller. NBW
acknowledges the financial support of the Commission of the European
Community (ERB CHRX CT-930 351). PN thanks the DFG for financial
support (Heisenberg foundation)


\begin{table}
\begin{center}
\caption{The fourth order cumulant ratio for the tricritical fixed point
operator distributions.}

\begin{tabular}{|cc|}
$U_L^{\cal M}$   & $0.348(3)$  \\
$U_L^{\cal D}$   & $0.574(2)$  \\
$U_L^{\cal E}$   & $0.003(3)$ \\
\end{tabular}
\label{tab:ULs}
\end{center}

\end{table}

\begin{figure}[h]

\caption{{\bf (a)} Schematic phase diagram of the spin fluid in the
$T$--$\rho$ plane. {\bf (b)} Schematic phase diagram in the $\mu$--$T$ plane
showing the directions of the relevant scaling field $g$ and $\lambda$.}

\label{fig:pdschem}
\end{figure}

\begin{figure}[h]

\caption{The measured cumulant ratio $U_L^{\cal D}$ for the 2d
Blume-Capel model along the first order line and its analytic
extension, determined according to the procedure described in the
text.}

\label{fig:BC_UL}
\end{figure}

\begin{figure}[h]

\caption{The line of first order transitions and its finite-size
analytical extension in the $\mu$--$T$ plane, obtained according to
the procedure described in the text.}

\label{fig:mu_T}
\end{figure}

\begin{figure}[h]
\caption{The measured cumulant ratio $U_L^{\cal D}$ for the 2d spin fluid model
along the first order line and its analytic extension
determined according to the procedure described in the text.}
\label{fig:fl_UL}
\end{figure}

\begin{figure}[h]

\caption{Typical particle/spin configurations of the $l=20$ spin fluid
near tricriticality. Spins values of $+1$ are denoted by filled circles,
and spin values $-1$ by unfilled circles.}

\label{fig:config}
\end{figure}

\begin{figure}[h]

\caption{The scaling operator distributions for the 2d spin fluid at
the designated tricritical parameters for each of the three system
sizes $l=12,16,20$. {\bf (a)} $\tilde{p}_L^\star({\cal M})$, {\bf
(b)}, $\tilde{p}_L^\star({\cal D})$ {\bf (c)} $\tilde{p}_L^\star({\cal
E})$.  Also shown for comparison are the corresponding distribution
measured for the tricritical $L=40$ 2d Blume-Capel model. All
distributions are expressed in terms of the scaling variable
$a_i^{-1}L^{d-y_i}({\cal O}-{\cal O}_c)$ and are scaled to unit norm
and variance. Statistical errors do not exceed the symbol sizes.}

\label{fig:collapse}
\end{figure}

\newpage
\begin{figure}[h]

\caption{The measured tricritical point forms of: {\bf (a)}. The
magnetisation distribution $p_L(m)$ (the data points have been
suppressed for clarity). {\bf (b)} The density distribution
$p_L(\rho)$. {\bf (c)} The energy density distribution $p_L(u)$. }

\label{fig:flbare}
\end{figure}


\begin{thebibliography}{50}

\bibitem{LAWRIE} For a general review of tricritical phenomena see
I. D. Lawrie and S. Sarbach, in {\em Phase transitions and critical
phenomena} edited by C. Domb and J.L. Lebowitz (Academic London,
1984), Vol. 8.

\bibitem{SAUL} D.M. Saul, M. W. Ortis and D. Stauffer, Phys. Rev. {\bf
B9}, 4964 (1974).

\bibitem{FURMAN} D. Furman, S. Dattagupta and R.B. Griffiths, Phys.
Rev. {\bf B15}, 441 (1977).

\bibitem{OLIVEIRA} S. M. de Oliveira, P.M.C. de Oliveira and F.C. de
S\'{a} Barreto, J. Stat. Phys. {\bf 78}, 1619 (1995).

\bibitem{BAKCHICH} A. Bakchich, A. N. Benyoussef and M. Touzani,
Physica {\bf A186}, 524 (1992).

\bibitem{BURKHARDT} T.W. Burkhardt and H.J.F. Knops, Phys. Rev. {\bf
B15}, 1602 (1977)

\bibitem{BERKER}A.N. Berker and H.J.F. Knops, Phys. Rev. B {\bf
15}, 1602 (1977).

\bibitem{YEOMANS} J. M. Yeomans and M.E. Fisher, Phys. Rev. B{\bf 24},
2825 (1981)

\bibitem{BEALE} P.D. Beale, Phys. Rev. {\bf 33}, 1717 (1986)

\bibitem{HERRMANN} H.J. Herrmann, Phys. Lett {\bf A100}, 256 (1984)

\bibitem{RIKVOLD} P.A. Rikvold, W. Kinzel, J.D. Gunton and K. Kaski,
Phys. Rev. B{\bf 28}, 2686 (1983)

\bibitem{ALCARAZ} F.C. Alcaraz,J.R. Drugowich de Felicio, R.
K\"{o}berle and J.F. Stilck, Phys. Rev. {\bf B32}, 7469 (1985)

\bibitem{AYAT} K.L. Ayat and C.M. Care, J. Magnet. and
Magn. Mater. {\bf 127}, L20 (1993)

\bibitem{KIMEL}J.D. Kimel, S. Black, P. Carter and Y.-L. Wang, Phys.
Rev. {\bf B35}, 3347 (1987).

\bibitem{LANDAU} D.P. Landau and R.H. Swendsen, Phys. Rev. {\bf B33},
7700 (1986)

\bibitem{LANDAU1} D.P. Landau and R.H. Swendsen, Phys. Rev. Lett. {\bf
46}, 1437 (1981)

\bibitem{HONDA} Y. Honda, Phys. Lett {\bf A184}, 74 (1993)

\bibitem{STRATT} See e.g. R.M. Stratt, J. Chem. Phys. {\bf 80}, 5764
(1988); D. Marx, P. Nielaba and K. Binder, Phys. Rev. {\bf
B47}, 7788 (1993); P. de Smedt, P. Nielaba, J.L. Lebowitz, J. Talbot
and L. Dooms, Phys. Rev. {\bf A38} 1381 (1988) and references therein.

\bibitem{KADANOFF} L. P. Kadanoff in {\em Phase transitions and
critical phenomena} edited by C. Domb and M. S. Green
(Academic New York, 1976), Vol 5A, pp 1-34.

\bibitem{WILDING3} N.B. Wilding and A.D. Bruce,
J. Phys. Condens. Matter {\bf 4}, 3087 (1992); A.D. Bruce and
N.B. Wilding, Phys. Rev. Lett. {\bf 68}, 193 (1992).

\bibitem{WILDING4} N.B. Wilding, Phys. Rev. {\bf E} (in press) (1995).

\bibitem{WILDING1} N.B. Wilding, Z. Phys. {\bf B93}, 119 (1993).

\bibitem{WILDING2} N.B. Wilding and M. M\"{u}ller, J. Chem. Phys. {\bf
102}, 2562 (1995).

\bibitem{MUELLER} M. M\"{u}ller and N.B. Wilding, Phys. Rev. {\bf
E51}, 2079 (1995).

\bibitem{GRIFF} R.B. Griffiths, Phys. Rev. {\bf B7}, 545 (1973)

\bibitem{WEGNER} F.J. Wegner, Phys. Rev. {\bf B5}, 4529 (1972).

\bibitem{REHR} J.J. Rehr and N.D. Mermin, Phys. Rev. A{\bf 8}, 472
(1973).

\bibitem{PRIVMAN} For a review of finite-size scaling methods, see
V. Privman (ed.) {\em Finite size scaling and numerical simulation of
statistical systems} (World Scientific, Singapore) (1990).

\bibitem{NOTE1} It should be noted that for large values of the
argument $a_1L^{y_1}g$ a crossover occurs to standard Ising critical
behaviour on the critical line. Such crossover phenomena can also be
explicitly incorporated within the finite-size scaling framework, see
eg. K. Binder and H.-P. Deutsch, Europhys. Lett. {\bf 18}, 667 (1992).

\bibitem{AD3} D.J. Adams, Mol. Phys. {\bf 29}, 307 (1975).

\bibitem{ALLEN} M.P. Allen and D.J. Tildesley {\em Computer simulation of
liquids}
Oxford University Press (1987).

\bibitem{FERRENBERG} A.M. Ferrenberg and R.H. Swendsen,
Phys. Rev. Lett. {\bf 61}, 2635 (1988); {\it ibid} {\bf 63}, 1195
(1989).

\bibitem{BINDER2} K. Binder, Z. Phys. {\bf B43}, 119 (1981).


\bibitem{CRAMER} H. Cram\'{e}r {\it Mathematical Methods of Statistics}
(Princeton NJ : Princeton University Press 1946)


\bibitem{LAJZEROWICZ} J. Lajzerowicz amd J. Sivardi\`{e}re, Phys. Rev.
{\bf B11} 2079 (1975).

\bibitem{NEINHUIS} B. Neinhuis, A.N. Berker, E.K. Reidel and M.
Schick, Phys. Rev. Lett. {\bf 43}, 737 (1979).

\bibitem{NIJS} M.P.M. den Nijs, J. Phys. A{\bf 12}, 1857 (1979).

\bibitem{PEARSON} R.B. Pearson, Phys. Rev. B{\bf 22}, 2579 (1980).

\bibitem{NICOLAIDES} D. Nicolaides and A.D. Bruce, J. Phys. A. {\bf
21}, 233 (1988).


\end{thebibliography}
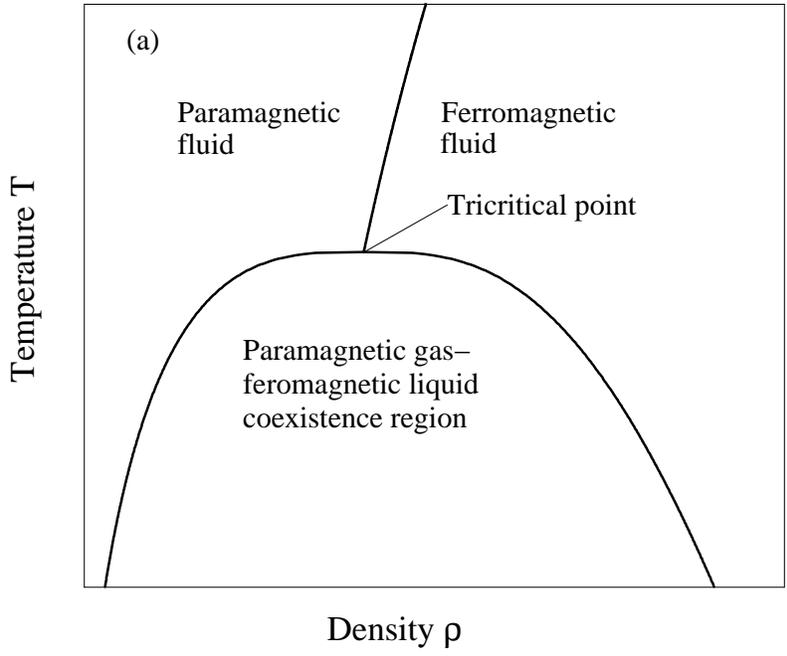
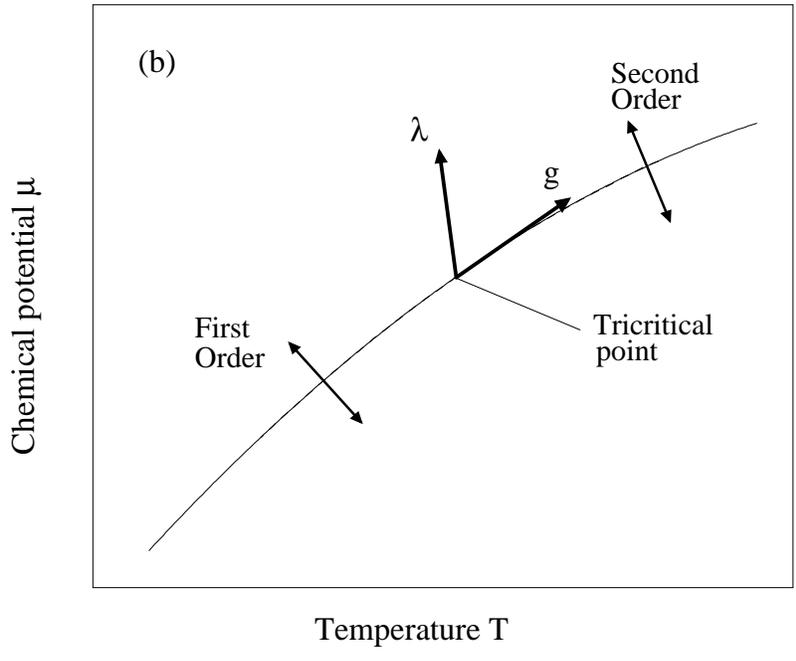
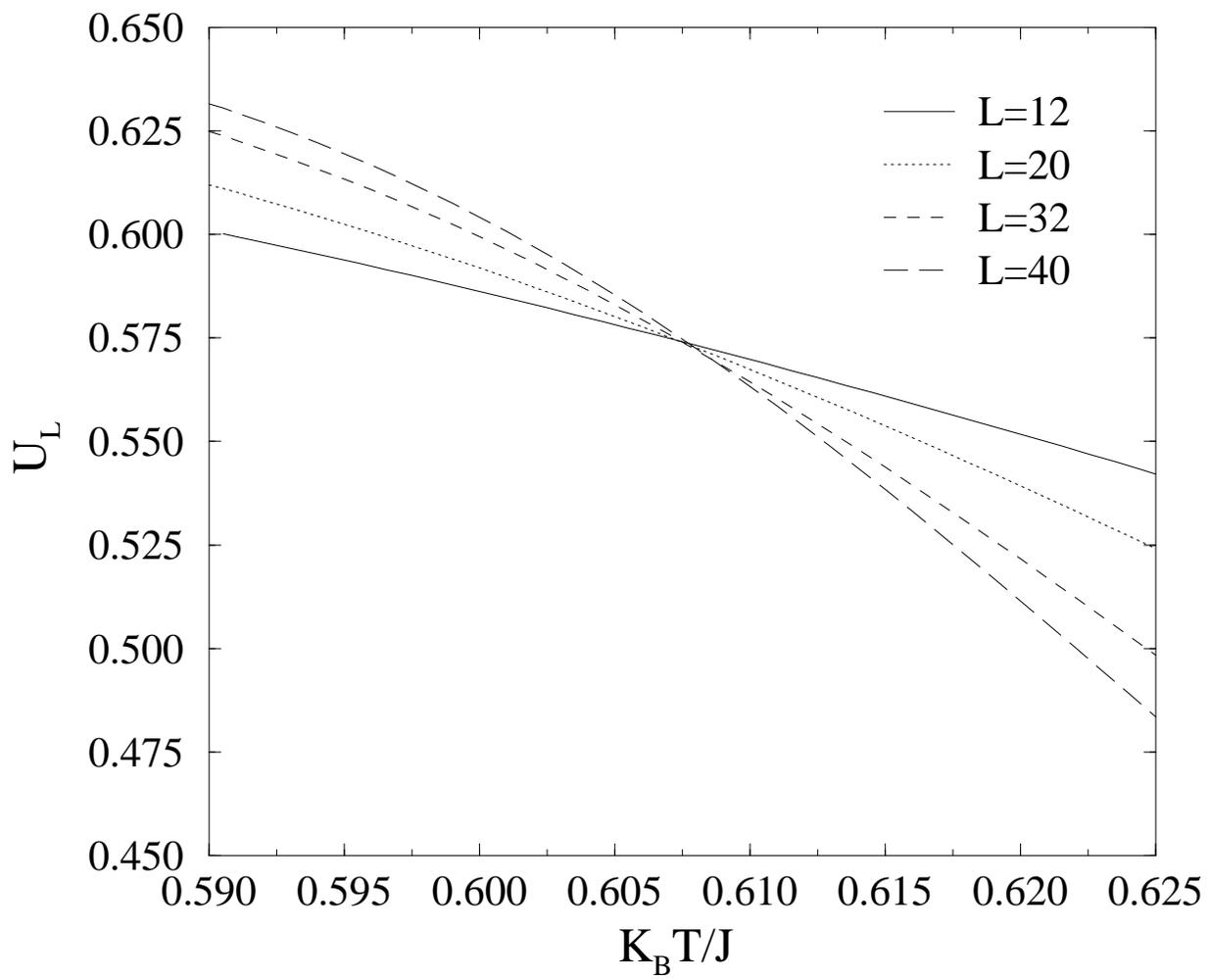
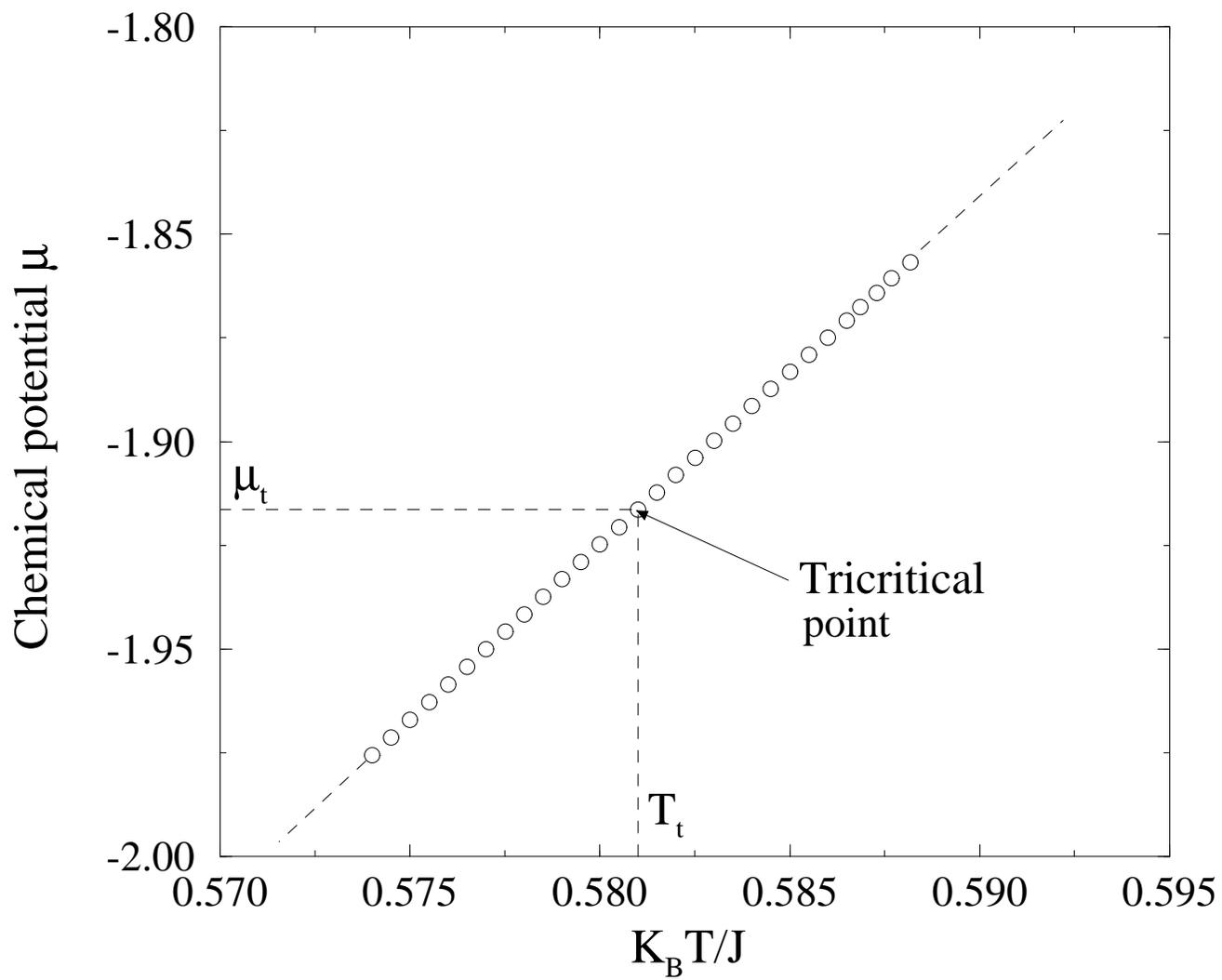
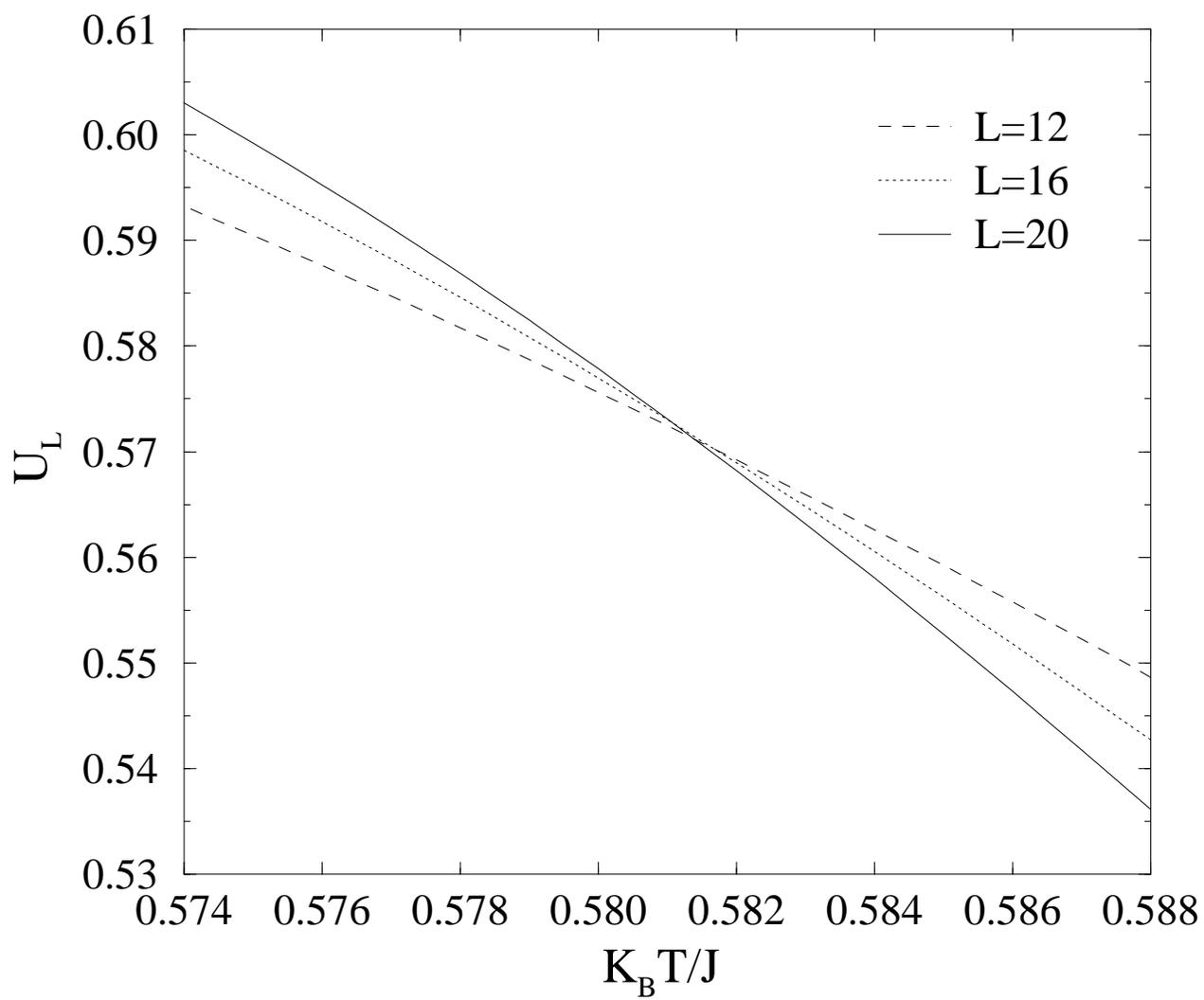
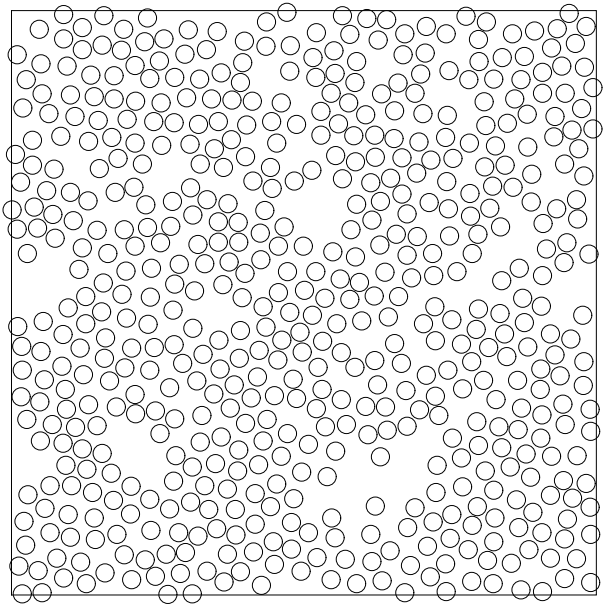
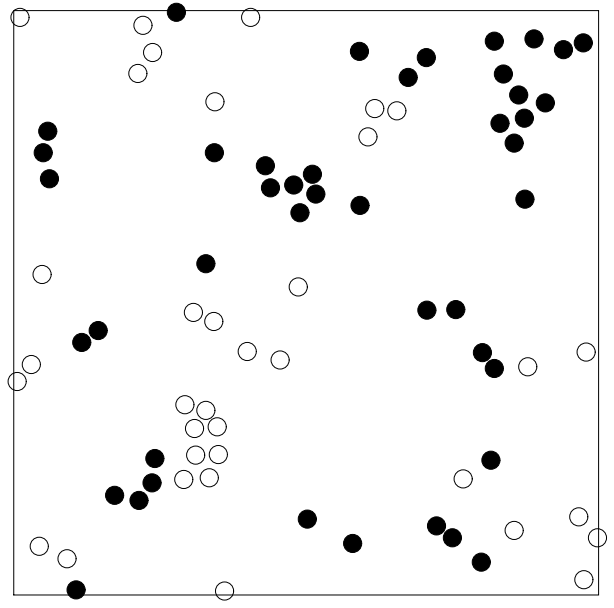
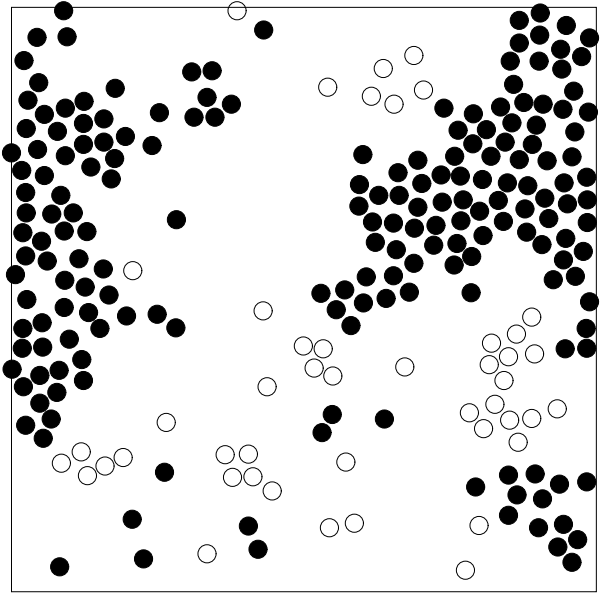
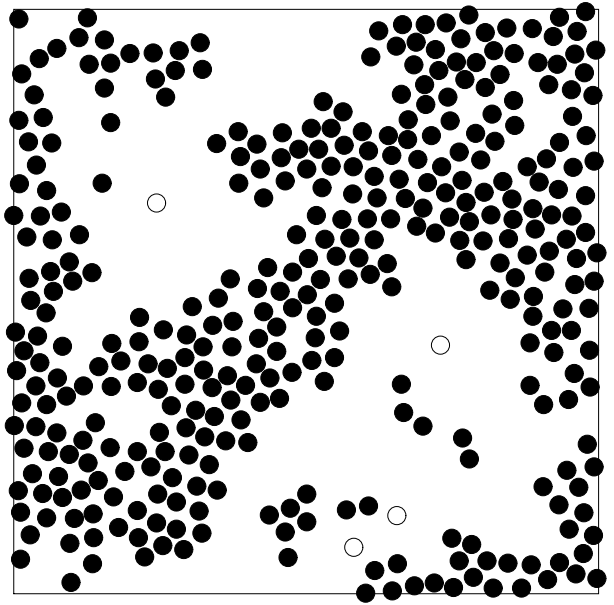
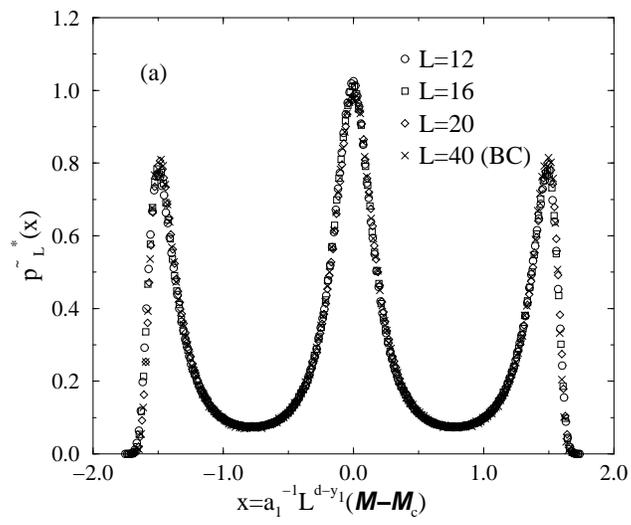
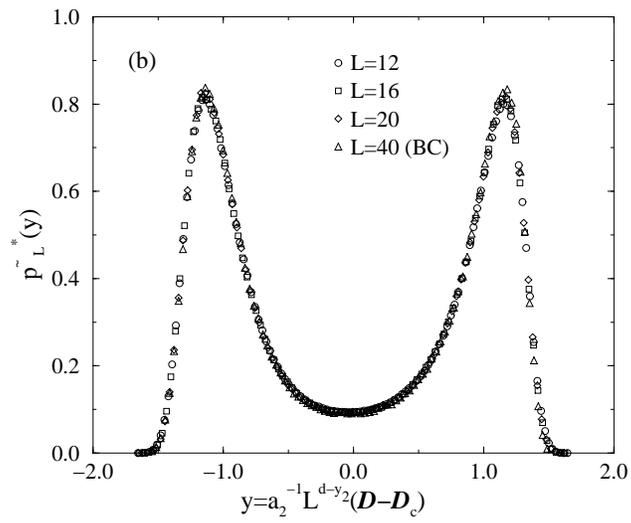
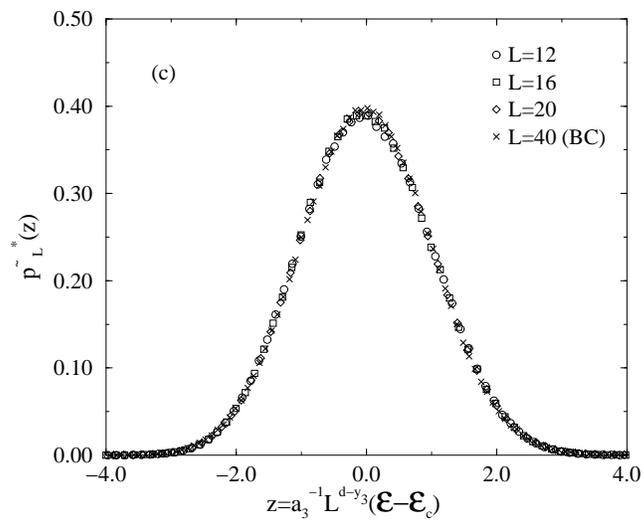
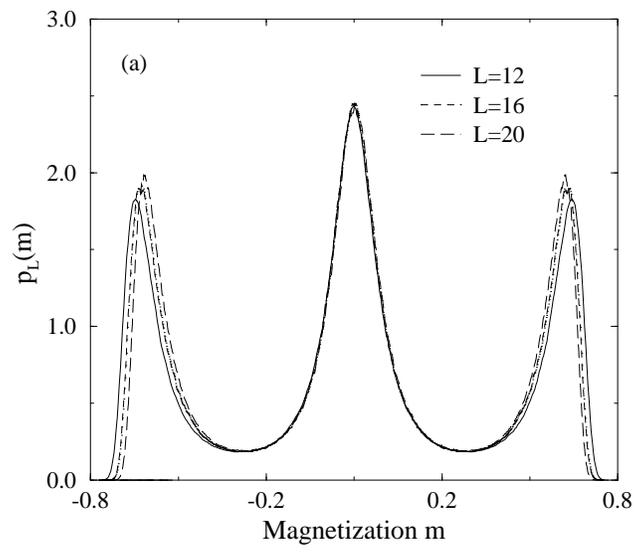
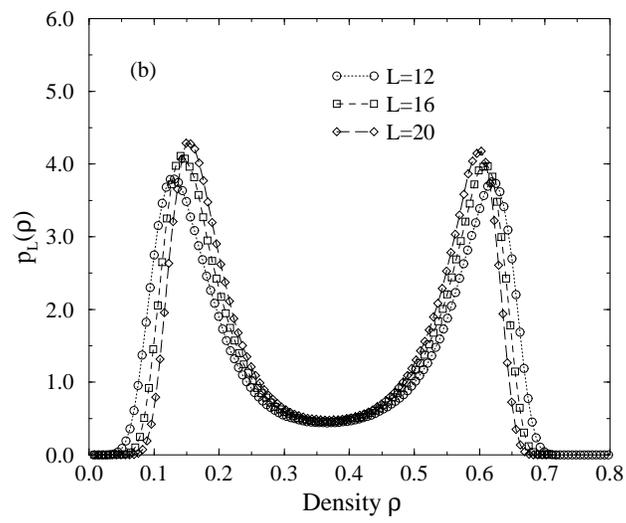
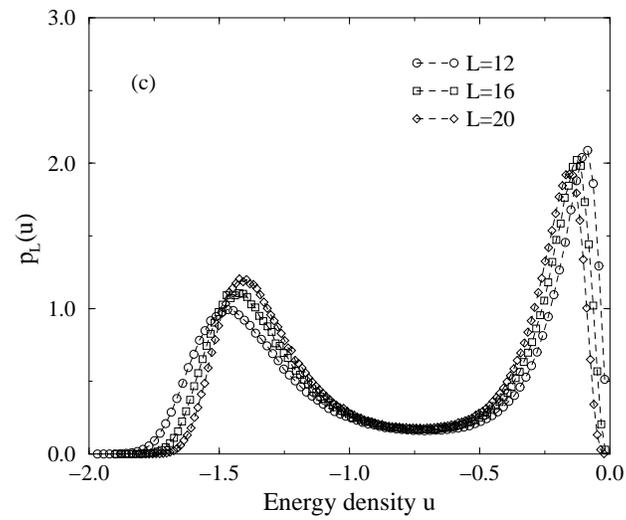
\end{document}